%% file: main.tex
\documentclass[runningheads]{llncs}
\usepackage[T1]{fontenc}
\usepackage{amsmath}
\usepackage{graphicx}
\usepackage{forest}
\usepackage{amssymb}
\renewcommand{\squareforqed}{$\square$}
\let\llncsendproof\endproof
\renewcommand{\endproof}{\qed\llncsendproof}

\usepackage{float}
\usepackage{algorithm}
\usepackage{algpseudocode}

\newfloat{dataset}{t}{lod}
\floatname{dataset}{Dataset element}
\usepackage{tabularx}
\usepackage{booktabs}
\newcolumntype{Y}{>{\centering\arraybackslash}X} 
\usepackage{comment}
\usepackage{pgfkeys}
\usepackage{tikz}
\usetikzlibrary{
  positioning,
  shapes.geometric,
  arrows.meta,
  backgrounds,
  fit
}

\usepackage{newfloat}
\usepackage{listings}
\floatstyle{ruled}
\newfloat{listing}{tb}{lst}{}
\floatname{listing}{Listing}

\begin{document}
\title{A Neuro-Symbolic Approach to Strategy Synthesis for Strategic Logics}
%
%
\author{Marco Aruta\inst{1} \and Vadim Malvone \inst{2} \and Aniello Murano \inst{1}\and Domenico Parente \inst{3} \and Luca Rizzuti \inst{3}}
\authorrunning{M. Aruta et al.}
%
\institute{University of Naples Federico II, Naples, Italy \and LTCI, Télécom Paris, Institut Polytechnique de Paris, Palaiseau, France \and Università degli Studi di Salerno, Salerno, Italy }
\maketitle              
\begin{abstract}
Reasoning about what agents can achieve through strategic interaction is a core challenge in Multi-Agent Systems (MAS). Logics for strategic ability, such as ATL, provide rigorous methods, but their adoption is often hindered by the computational cost of strategy synthesis. We introduce a neuro-symbolic framework that integrates large language models (LLMs) into the model-checking pipeline for MAS. The LLM acts as a strategy-generation oracle, proposing candidate strategies that are then formally validated by a standard MAS model checker. This generate-and-certify architecture uses LLM guidance to navigate large combinatorial strategy spaces while preserving formal soundness: generated strategies are accepted only when certified by the verifier. We instantiate the framework for bounded strategic reasoning in NatATL and introduce the first NatATL strategy-synthesis dataset, consisting of 4,211 instances. Experiments with an open-weight Qwen3-32B model show that our certified pipeline achieves 92\% accuracy on strategy-synthesis outcomes.
\end{abstract}

\input{sections/01-intro}
\input{sections/02-preliminaries}

\input{sections/03-ds}
\input{sections/04-framework}
\input{sections/05-Evaluation}
\input{sections/06-Conclusions}

%
%
%
\bibliographystyle{splncs04}
\bibliography{biblio}

\end{document}

%% file: sections/01-intro.tex
\section{Introduction}
Strategic reasoning in Multi-Agent Systems (MAS) is a central topic in Artificial Intelligence, with applications ranging from autonomous systems and robotics to security, economics, and distributed control~\cite{shoham2008multiagent,wooldridge2009introduction}. Temporal logics for strategies, such as Alternating-time Temporal Logic (ATL) and related formalisms, provide a rigorous framework to reason about the abilities of agents and coalitions to enforce temporal objectives under strategic interaction~\cite{alur2002alternating,aagotnes2006action,chatterjee2010strategy,DBLP:journals/tocl/MogaveroMPV14}. Among these, Natural Alternating-time Temporal Logic (NatATL) was introduced to capture \emph{natural strategic ability}~\cite{JMM19a}, where agents are constrained to strategies that are cognitively simple, interpretable, and aligned with \emph{bounded} rationality assumptions. Real-world agents operate under computational and cognitive limits: strategies with unbounded memory, although theoretically expressive, are often impossible to implement in practice. NatATL formalizes this insight by restricting attention to strategies that can realistically be executed by boundedly rational agents, thereby bridging the gap between formal strategic reasoning and implementable strategic behavior.

Despite its conceptual relevance, NatATL faces two major obstacles that limit its practical adoption. First, existing model-checking procedures rely on explicit enumeration of admissible natural strategies, leading to an exponential blow-up in the strategy space. Approaches based on exhaustive generation~\cite{kaminski2025natstv} become quickly infeasible even for small models or coalitions. The practical bottleneck is therefore not only checking a fixed candidate strategy, but discovering a promising bounded natural strategy in a vast combinatorial space. Second, unlike mainstream learning-based or planning-based approaches, NatATL lacks structured, labelled datasets that could support data-driven methods for strategy synthesis or approximation. This absence of benchmarks further hinders the exploration of alternative computational paradigms beyond brute-force reasoning. Recent advances in Large Language Models (LLMs) offer a promising direction to address these challenges. LLMs have shown strong capabilities in synthesizing structured symbolic artifacts from high-level specifications, implicitly navigating large combinatorial spaces while favoring concise and human-aligned representations~\cite{austin2021program,pan2023logic,tantakoun2025llms}. However, LLM-only synthesis remains unreliable in the sense required by formal verification: outputs can be subtly incorrect, semantically inconsistent with the underlying model, or brittle under adversarial counter-strategies. In safety-critical and formally verified settings, this lack of intrinsic reliability is unacceptable: what is needed is not merely a plausible strategy, but a \emph{certified} one.


\paragraph{Our Contributions.}
We propose a neuro-symbolic framework for certified strategy synthesis in NatATL. The main contributions of this paper are as follows:

\begin{itemize}
    \item \textbf{LLM-guided certified strategy synthesis.}
    We introduce a generate-and-certify framework in which an LLM acts as a strategy-generation oracle for NatATL. The LLM proposes candidate natural strategies directly from the multi-agent model and the NatATL specification, while a standard NatATL model checker formally validates each candidate. Thus, the framework exploits LLM guidance to navigate large combinatorial strategy spaces while preserving soundness: only verifier-certified strategies are accepted.

    \item \textbf{The first NatATL strategy-synthesis dataset.}
    We construct the first labelled dataset for NatATL strategy synthesis, consisting of 4,211 instances of concurrent multi-agent models, NatATL specifications, and expert-validated natural strategies. The dataset is built through a human-in-the-loop validation process and is designed to support reproducible evaluation of strategy-generation oracles and future learning-based approaches.

    \item \textbf{An open-weight and reproducible implementation.}
    We instantiate the framework using an open-weight Qwen3-32B model integrated with the VITAMIN NatATL verifier \cite{DBLP:conf/ifaamas/0001M25}. The resulting pipeline supports end-to-end strategy generation, syntactic validation, formal certification, and verifier-guided refinement, providing a reproducible neuro-symbolic toolchain for NatATL reasoning.

    \item \textbf{Empirical evidence of effectiveness and scalability.}
    We evaluate the framework on the proposed benchmark and show that the certified pipeline achieves 92\% accuracy on strategy-synthesis outcomes. The experiments also show that our approach mitigates the explicit natural-strategy enumeration bottleneck, scaling to configurations with up to 50 states, coalitions of 11 agents, and complexity bounds up to \(k=100\), while accepting only formally certified positive results.
\end{itemize}

\paragraph{Related Works.}
Logics for strategic reasoning in Multi-Agent Systems have been extensively studied as a formal foundation for analyzing the abilities of agents and coalitions under strategic interaction~\cite{van2005logic,van2008multi}. Alternating-time Temporal Logic (ATL)~\cite{alur2002alternating} and its numerous extensions (\cite{HP06,CL07a,jamroga2008temporal,10.5555/1838206.1838274,JamrogaKKP20,bulling2022combining,DBLP:conf/prima/FerrandoLMM24,catta2024resource,catta2024obstruction,DBLP:conf/ifaamas/JamrogaKPPS25}) constitute a well-established framework, with model-checking techniques that have been refined to address issues such as imperfect information, memory bounds, and resource constraints~\cite{DBLP:journals/tocl/MogaveroMPV14,DBLP:conf/tacas/ChenFKPS13,DBLP:conf/aaai/HuangM14,DBLP:conf/aaai/CermakLM15,DBLP:conf/atva/GutierrezNPW18,DBLP:journals/sttt/LomuscioQR17,JMM19b}. A long line of work has investigated the trade-off between expressiveness and computational complexity by restricting strategies, for instance to memoryless, bounded-memory, or resource-aware variants \cite{bulling2010model,dima2011model,berthon2018decidability,DBLP:conf/kr/AminofGLMR21,BJMM24}. In this landscape, Natural Alternating-time Temporal Logic (NatATL) was introduced to capture \emph{natural strategic ability}, enforcing strategies that are cognitively simple and human-interpretable~\cite{JMM19a,JMM19b}. Subsequent developments explored natural strategies in applied domains and enriched settings, including auctions, voting protocols, and quantitative or fuzzy extensions~\cite{jamroga2020natural,belardinelli2022reasoning,berthon2024natural,aruta2025natural}. Despite these advances, existing NatATL verification procedures fundamentally rely on explicit enumeration of admissible natural strategies, which leads to severe scalability limitations as the complexity bound or coalition size increases~\cite{kaminski2025natstv}. In parallel, several approaches have proposed symbolic, abstraction-based, or reduction techniques to mitigate state-space explosion in strategic logics~\cite{DBLP:conf/cav/CermakLMM14,DBLP:conf/atal/KurpiewskiPJK21}, yet they do not address the combinatorial blow-up inherent in natural strategy generation. More recently, large language models (LLMs) have demonstrated strong capabilities in synthesizing symbolic artifacts such as programs, logical rules, and proofs from high-level specifications~\cite{austin2021program,li2022competition,pan2023logic,tantakoun2025llms}.
Related work has explored the use of LLMs for formal reasoning tasks, including mathematical problem solving, program synthesis, and neural theorem proving~\cite{yang2023large,li2024neuro,liu2025logical}. However, these approaches typically lack formal semantic guarantees and are not grounded in logics of strategic ability. In parallel, LLMs have been employed in agent-based and planning-oriented settings~\cite{yao2022react,shinn2023reflexion}, but without a formal treatment of strategic interaction, bounded rationality, or temporal objectives. To the best our knowledge, no prior work integrates LLM-guided strategy synthesis with sound NatATL model checking, nor provides datasets to support such neuro-symbolic approaches. Our work combines LLM-based natural strategy synthesis with formal verification and introduces the first expert-validated dataset for NatATL.

\paragraph{Outline.}Section~2 introduces the necessary preliminaries on natural strategies and NatATL verification. Section~3 presents the dataset. Section~4 describes the proposed LLM-guided verification framework. Section~5 reports the experimental evaluation. Section~6 concludes the paper.

%% file: sections/02-preliminaries.tex
\section{Preliminaries}

We briefly recall the main notions underlying Natural Alternating-time Temporal Logic (NatATL) and its model-checking problem. We assume familiarity with standard concepts from temporal logic and multi-agent systems, and focus on the aspects that are essential for our framework.

\paragraph{Concurrent Game Structures.}
A \emph{Concurrent Game Structure} (CGS) is the standard semantic model for logics of strategic ability. Formally, a CGS is a tuple
$G = (Ag, Ap, \{Act_a\}_{a \in Ag}, St, s_0, d, \delta, \ell)$,
where $Ag$ is a finite set of agents, $Ap$ a finite set of atomic propositions, $Act_a$ the finite set of actions available to agent $a$, $St$ a finite set of states with initial state $s_0$, $d(a,s) \subseteq Act_a$ the non-empty set of actions available to agent $a$ at state $s$, $\delta$ the transition function mapping states and joint actions to successor states, and $\ell : St \to 2^{Ap}$ a labeling function.
A \emph{joint action} is a tuple selecting one action per agent, and the evolution of the system is determined by the collective choices of all agents.

\paragraph{Natural Strategies.}
Unlike classical ATL, which allows arbitrary (possibly highly complex) strategies, NatATL restricts attention to \emph{natural strategies}, intended to model human-like, interpretable decision rules. A natural memoryless strategy for an agent $a$ is given by a finite, ordered list of guarded actions
\[
s_a = \langle (\varphi_1,\alpha_1), \ldots, (\varphi_n,\alpha_n) \rangle,
\]
where each guard $\varphi_i$ is a propositional formula over $Ap$, and $\alpha_i$ is an action available to agent $a$ in every state satisfying $\varphi_i$. The last rule is required to have guard $\top$, ensuring that the strategy is total.
At a given state, the agent executes the action associated with the first guard that holds. A \emph{collective natural strategy} for a coalition $A \subseteq Ag$ is a tuple of individual strategies, one for each agent in $A$. The outcome of such a strategy from a state $s$ is defined as the set of all paths that are compatible with the prescribed actions of the coalition, while the agents outside the coalition are unrestricted.

\paragraph{Strategy Complexity.}
The defining feature of NatATL is that natural strategies are subject to a syntactic \emph{complexity bound}. Complexity is measured as the size of the propositional guards (e.g., the number of symbols or atomic propositions occurring in them). A collective strategy is admissible if the sum of the complexities of its component strategies does not exceed a given bound $k$. This formalizes bounded rationality by limiting the descriptive effort required to specify a strategy.

\paragraph{Model Checking Problem.}
Given a formula of the form
$\langle\!\langle A \rangle\!\rangle^{\leq k}\psi$, the NatATL
model-checking problem asks whether there exists a collective natural strategy
$s_A$ for the coalition $A$ such that $\mathrm{compl}(s_A)\leq k$ and every
path induced by $s_A$ satisfies the temporal objective $\psi$. Existing
procedures follow this semantics directly: they search over admissible bounded
natural strategies and check, for each candidate, whether the induced outcomes
satisfy the temporal objective~\cite{JMM19a}. Thus, the key scalability barrier
is the explicit construction of the candidate strategy space.

This bottleneck comes from the representation of natural strategies. In the
memoryless case, a natural strategy is an ordered list of guarded actions, where
each guard is a Boolean condition over atomic propositions. Therefore, the
verifier must consider possible guards, possible actions, possible rule
orderings, and possible ways of distributing the global complexity budget across
the agents in the coalition. For strategies of complexity at most $k$, the
number of memoryless candidates is bounded by
\[
N_{\mathsf{nr}}(k)
 =
O\!\left((|Prop|+|Bool|)^{k^2}|Act|^k\right).
\]
This term is already exponential in the bound $k$. Consequently, even when
model checking is polynomial in the model size for fixed $k$, the hidden
constant contains the bounded-strategy enumeration factor.

The situation becomes more demanding for natural strategies with recall. In
that setting, guards are regular expressions over Boolean propositional
formulas. The corresponding candidate space is bounded by
\[
N_{\mathsf{nR}}(k)
 =
O\!\left((|Prop|+|Bool|+|Con|)^{k^2}|Act|^k\right),
\]
where $Con$ denotes the regular-expression constructors. Moreover, checking a
fixed recall-based strategy requires reasoning over a bounded unfolding induced
by the strategy and the automata associated with its guards. Table~\ref{tab:complexity}
summarizes the relevant NatATL settings.

\begin{table}[t]
\centering
\scriptsize
\setlength{\tabcolsep}{2.2pt}
\renewcommand{\arraystretch}{1.08}
\begin{tabular}{lcc}
\hline
\textbf{Setting} & \textbf{MC bound} & \textbf{Search bottleneck} \\
\hline
Mem., fixed $k$
&
in P
&
$N_{\mathsf{nr}}(k)$ \\
Mem., var. $k$
&
$\Delta^P_2$-complete
&
exp. in $k$ \\
Recall, fixed $k$
&
in $\Delta^P_2$
&
$N_{\mathsf{nR}}(k)$ + unfold. \\
Recall, var. $k$
&
in PSPACE
&
exp. in $k$ + unfold. \\
\hline
\end{tabular}
\caption{NatATL model-checking bounds and natural-strategy generation
bottlenecks. Here
$N_{\mathsf{nr}}(k)=O((|Prop|+|Bool|)^{k^2}|Act|^k)$ for memoryless
strategies, while
$N_{\mathsf{nR}}(k)=O((|Prop|+|Bool|+|Con|)^{k^2}|Act|^k)$ for recall-based
strategies.}
\label{tab:complexity}
\end{table}
From an algorithmic perspective, the table shows that the practical bottleneck
of NatATL model checking is the search for a bounded natural strategy witness.
Once a memoryless candidate strategy is fixed, the model can be pruned according
to the actions prescribed by the strategy and the temporal objective can be
checked on the resulting structure. The expensive step is finding the candidate
in the first place. This is precisely the step that dominates existing
explicit-generation procedures as the complexity bound, the number of actions,
or the coalition size increases.

%% file: sections/03-ds.tex
\section{Dataset}
\label{ds}

    The lack of structured benchmarks for NatATL limits the systematic evaluation of data-driven and neuro-symbolic approaches to natural strategy synthesis. Existing NatATL verification procedures are primarily designed around exhaustive generation and checking of bounded natural strategies~\cite{kaminski2025natstv,aruta2025s4h}. While this yields a sound verification procedure, it does not provide a shared collection of NatATL instances on which alternative strategy-generation mechanisms can be evaluated in a reproducible way. 
    
    To address this gap, we introduce a NatATL benchmark dataset consisting of 4211 concurrent multi-agent models paired with NatATL specifications. Each instance describes a finite concurrent game structure, a strategic coalition, a complexity bound, and a NatATL objective to be checked. The dataset is designed as an evaluation benchmark for strategy-synthesis oracles: given a model and a formula, an external synthesizer, such as an LLM, proposes a natural strategy, which is then certified or rejected by the NatATL verifier. The instances are intentionally compact enough to be serialized and processed by an LLM within its context window, while stressing the main sources of combinatorial blow-up in NatATL verification: coalition size, branching over joint actions, and the complexity bound on natural strategies. This reflects the observation, discussed in the preliminaries, that the main scalability barrier is not only the number of states, but the explosion of admissible guarded strategies as the coalition and the bound grow~\cite{aruta2024model,kaminski2025natstv}.

\subsection{Generation Pipeline}

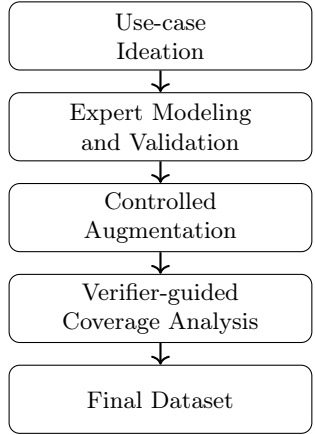
\begin{figure}[t]
\centering
\begin{tikzpicture}[
    node distance=1.2cm,
    every node/.style={draw, rectangle, rounded corners, align=center, minimum width=4.0cm, minimum height=0.9cm},
    arrow/.style={->, thick}
]

\node (ideation) {Use-case\\Ideation};
\node (filter) [below of=ideation] {Expert Modeling\\and Validation};
\node (augmentation) [below of=filter] {Controlled\\Augmentation};
\node (coverage) [below of=augmentation] {Verifier-guided\\Coverage Analysis};
\node (dataset) [below of=coverage] {Final Dataset};

\draw[arrow] (ideation) -- (filter);
\draw[arrow] (filter) -- (augmentation);
\draw[arrow] (augmentation) -- (coverage);
\draw[arrow] (coverage) -- (dataset);

\end{tikzpicture}
\caption{Pipeline for NatATL benchmark construction.}
\label{fig:dataset_pipeline}
\end{figure}

The dataset was constructed through a multi-stage process combining use-case ideation, expert modeling, automatic consistency checking, controlled augmentation, and verifier-guided coverage analysis. Figure~\ref{fig:dataset_pipeline} summarizes the overall pipeline.

The first stage focused on identifying application domains and interaction patterns suitable for natural strategic reasoning. A closed-source LLM (i.e., Gemini 3.1 Pro) was used only as a high-level ideation aid to suggest candidate use cases, including access control, robotics, logistics, traffic management, medical operations, industrial control, and cyber-physical monitoring. These suggestions were not used directly as formal data. Instead, five domain experts in strategic logics, multi-agent systems, and formal verification manually filtered the proposed scenarios, discarding cases that were not naturally expressible as concurrent game structures or did not involve meaningful strategic interaction. Starting from the filtered use cases, the experts manually constructed an initial seed set of NatATL instances. Each instance was encoded as a finite concurrent game structure with explicit states, agents, available actions, transitions, initial state, atomic proposition labeling, coalition, and NatATL formula. During this phase, the experts aimed to cover different strategic patterns, including safety objectives, reachability objectives, coordinated multi-agent control, adversarial disruption, and cases where the coalition is intentionally unable to enforce the objective under the specified complexity bound. This design choice was essential to avoid a benchmark biased toward trivially satisfiable formulas.

All seed instances underwent structural validation before augmentation. We checked both automatically and through experts human-in-the-loop inspection that every transition specifies the required joint-action information, that each action appearing in a transition belongs to the corresponding agent's declared action set, that all states referenced by transitions and labelings are declared, that the initial state belongs to the state space, that all agents appearing in the coalition are declared in the model, and that all atomic propositions occurring in the NatATL formula occur in the model labeling. We also normalized Boolean connectives and state identifiers to ensure compatibility with the NatATL parser and the VITAMIN bridge.

\subsection{Controlled Augmentation}
The validated seed set was expanded through controlled augmentation. This stage was designed to increase structural and lexical diversity while preserving formal well-formedness. We used three families of augmentation operators.

The first family consists of semantics-preserving renamings. These transformations systematically rename states, agents, actions, and atomic propositions while updating all dependent fields consistently. For example, when an agent is renamed, the transformation updates the agent list, the action dictionary, the coalition, the NatATL formula, and all auxiliary fields used internally during validation. Similarly, action renaming propagates to both the declared action sets and the joint actions occurring in the transition relation. These transformations prevent the benchmark from overfitting to a small vocabulary of recurring names.

The second family consists of formula-level transformations. These include increasing the complexity bound, decreasing it without going below the admissible minimum bound, changing Boolean structure in controlled ways, modifying local negations, and altering coalition membership when the resulting formula remains syntactically valid. Unlike pure renaming, these transformations are not intended to preserve satisfiability in general. Their purpose is to generate harder negative or borderline instances by changing the strategic requirements imposed on the coalition. In particular, decreasing the bound can turn a previously feasible natural strategy into an infeasible one, while increasing the bound preserves feasibility whenever a strategy was already available under the smaller bound. All generated formulas are re-parsed after augmentation to ensure NatATL well-formedness.

The third family consists of model-level transformations. These include changing selected state labelings, redirecting or removing individual transitions, adding admissible transitions, introducing trap states, and adding a controlled disturbing agent whose actions may interfere with coalition progress. These operators are used conservatively and only when the resulting model remains structurally valid. The goal is not to add arbitrary noise, but to generate plausible variants in which strategic success or failure depends on adversarial choices, missing progress transitions, or additional environmental interference.

All augmented instances were subjected to the same structural checks as the seed instances. In addition, formulas with invalid complexity bounds were rejected or corrected according to the NatATL encoding used by our verifier. In particular, for non-empty coalitions we require the bound to be at least one, since a natural strategy contains at least one guarded rule, whereas the zero bound is only meaningful for degenerate encodings such as the empty-coalition universal-path case, which is not relevant to our synthesis objective.

\paragraph{Verifier-guided coverage analysis.}
After the first augmentation stage, we used the NatATL verification pipeline to inspect the semantic coverage of the generated instances, to avoid the dataset being dominated by a single outcome class. This is important because, for instance, an overwhelmingly negative benchmark would make it difficult to assess whether a synthesis oracle can both refrain from proposing impossible strategies and generate certifiable winning strategies when such strategies exist. Accordingly, we used verifier outcomes to identify underrepresented satisfiable and borderline regions of the instance space, and then applied an additional conservative augmentation round to improve coverage of these regions. This second augmentation round used only transformations that preserve, or monotonically relax, the underlying winning pattern. In particular, we allowed agent renaming, action renaming, reordering of transitions, reordering of declared actions, and increases of the complexity bound. We avoided transformations that may change satisfiability, such as changing the labeling, redirecting transitions, decreasing the bound, modifying the temporal objective, changing the coalition size, or adding new adversarial dynamics. This conservative policy preserves the strategic structure of the source instances while increasing lexical and syntactic diversity. The final dataset contains 4211 NatATL instances. To prevent inflated evaluation due to near-duplicate augmented variants, all experimental splits are performed at the level of augmentation families rather than individual JSON objects. Thus, instances derived from the same seed are assigned to the same split. This avoids parent-child leakage between development and evaluation data and ensures that reported results measure generalization to unseen strategic structures rather than memorization of renamed variants.

\paragraph{Data format and schema.}
Each dataset element is represented as a JSON object. The released benchmark focuses on the fields required by the strategy synthesis and verification pipeline. The main field is \textit{input}, which contains the concurrent game structure and the NatATL specification:
\textit{states},
\textit{agents},
\textit{actions},
\textit{transitions},
\textit{initial\_state},
\textit{labeling},
\textit{coalition},
\textit{formula\_natatl}.
The field \textit{states} lists the finite state space; \textit{agents} lists the agents of the model; \textit{actions} maps each agent to its admissible actions; \textit{transitions} encodes the transition relation through source states, joint actions, and target states; \textit{initial\_state} specifies the state from which verification starts; \textit{labeling} maps states to atomic propositions; \textit{coalition} identifies the strategic coalition; and \textit{formula\_natatl} stores the bounded NatATL objective. Additional metadata track identifiers, parent-child relations introduced by augmentation, augmentation type, and a natural language note describing the strategic objective.

%% file: sections/04-framework.tex
\section{LLM-Guided NatATL Verification Framework}
\label{sec:framework}

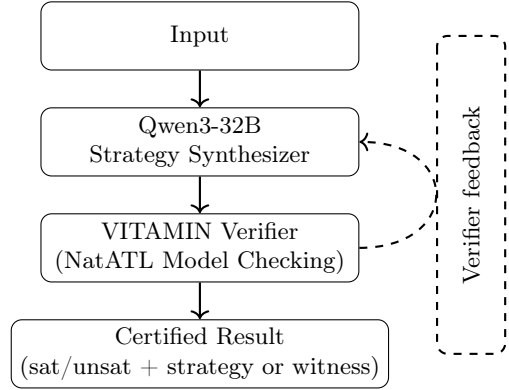
\begin{figure}[h]
\centering
\begin{tikzpicture}[
    node distance=1.4cm,
    every node/.style={
        draw,
        rectangle,
        rounded corners,
        align=center,
        minimum width=4.2cm,
        minimum height=0.9cm
    },
    arrow/.style={->, thick},
    feedback/.style={->, thick, dashed}
]

\node (data) {Input};
\node (llm) [below of=data] {Qwen3-32B\\Strategy Synthesizer};
\node (verifier) [below of=llm] {VITAMIN Verifier\\(NatATL Model Checking)};
\node (output) [below of=verifier] {Certified Result\\(sat/unsat + strategy or witness)};

\draw[arrow] (data) -- (llm);
\draw[arrow] (llm) -- (verifier);
\draw[arrow] (verifier) -- (output);

\draw[feedback]
(verifier.east) .. controls +(1.35,0) and +(1.35,0) ..
node[midway, sloped, below, font=\small] {Verifier feedback}
(llm.east);
\end{tikzpicture}

\caption{End-to-end tool pipeline. Given a NatATL instance, Qwen3-32B synthesizes candidate natural strategies, which are formally verified by the VITAMIN model checker. Unsatisfied candidates trigger an iterative verifier-guided refinement loop, while successful ones yield certified verification results.}
\label{fig:tool_pipeline}
\end{figure}

Figure~\ref{fig:tool_pipeline} summarizes the end-to-end workflow of our tool, which integrates an open-weight LLM, Qwen3-32B, with the VITAMIN NatATL verifier in a
verifier-in-the-loop synthesis setting. Each input instance consists of a concurrent game structure, a labeling function, a NatATL specification, a strategic coalition, and a complexity bound. The pipeline first converts this information into a structured prompt that exposes the relevant model components and specifies the required strategy format. The LLM then generates a candidate memoryless natural strategy. In our setting, a candidate strategy is represented as an ordered list of guarded actions for each agent in the coalition, together with a default rule ensuring totality. Before invoking the verifier, the generated output is parsed and checked against the expected schema. Outputs that are not valid JSON, refer to unknown agents, states, propositions, or actions, omit required default rules, or violate the declared strategy format are rejected before semantic verification. If the candidate passes these preliminary checks, it is submitted to VITAMIN. The verifier checks whether the strategy is well formed, respects the NatATL
complexity bound, prescribes only admissible actions, and satisfies the target objective under the NatATL semantics. When the strategy is winning, VITAMIN
returns a certified positive result together with the synthesized strategy as a witness. When the strategy is rejected, the verifier returns diagnostic
information, which can be translated into feedback for a bounded refinement round. The loop terminates when a strategy is certified, when the LLM fails to
produce a valid candidate, or when the retry budget is exhausted.

\paragraph{Prompt Design.}
The prompt was designed to make the LLM produce explicit memoryless natural strategies in a format that can be parsed and checked automatically. Each prompt contains the concurrent game structure, the coalition, the complexity bound, and the NatATL objective. The required output is a JSON object encoding, for each coalition agent, an ordered list of guarded actions together with a default rule. We initially evaluated both fine-tuning and one-shot prompting strategies using Qwen2.5-32B-4bit. In preliminary experiments, Qwen2.5 often failed to produce stable JSON outputs conforming to the required schema, independently of whether fine-tuning was used. We therefore switched to Qwen3-32B-4bit, whose reasoning mode produced substantially more stable structured outputs. Based on this behavior, we adopted a one-shot prompting strategy rather than a fine-tuning pipeline. Prompt development was performed on a development subset of 200 instances. At each iteration, we analyzed representative failures, including invalid JSON, incorrect interpretation of NatATL objectives, missing default rules, complexity-bound violations, and strategies rejected by the verifier. The prompt
was then refined to make the output schema more explicit and to clarify the semantics of guarded natural strategies. The refinement process stabilized after approximately six iterations. Finally, we introduced a compact markdown representation of the input model, derived from the original JSON encoding. This representation reduces prompt length and makes states, labels, actions, and transitions easier to inspect. In our development experiments, this conversion improved accuracy by approximately 3 percentage points, likely by reducing attention dilution caused by long JSON encodings of multi-agent systems.

\paragraph{Implementation Details.}
The pipeline is entirely implemented in Python and integrates Qwen3-32B with the VITAMIN model checker. All inference experiments were run on a Linux machine with
12 CPU cores, 64 GB of RAM, and an NVIDIA Quadro RTX A5000 GPU. We used the 4-bit quantized model \texttt{unsloth/Qwen3-32B-bnb-4bit} with reasoning mode enabled. VITAMIN is used as the verification backend for NatATL~\cite{DBLP:conf/ifaamas/0001M25}. This choice is motivated by its open-source implementation and by its modular architecture, which allows direct integration with Python and explicit inspection of strategy-evaluation components. In our pipeline, VITAMIN is responsible for checking whether a candidate strategy is syntactically valid, respects the NatATL complexity bound, prescribes only admissible actions, and satisfies the target formula on the given model. Before invoking the verifier, each LLM output is parsed and checked against the strategy schema. Invalid outputs are classified as parsing or schema failures. These include malformed JSON, unknown agents, unknown actions, malformed guards, missing default rules, and outputs that cannot be mapped to the coalition specified in the input formula. When allowed by the retry budget, such failures are converted into feedback for a subsequent LLM attempt. The same feedback mechanism is used when VITAMIN rejects a syntactically valid but losing strategy. In that case, the verifier diagnosis is translated into a natural-language message describing why the candidate failed, for example because a prescribed action is not admissible, a guard does not cover the required cases, or the induced outcomes violate the temporal objective. This feedback is then appended to the next query in the refinement loop.

\paragraph{Outcome Classification.}
\label{subsec:outcome-classification}

To evaluate the LLM-guided synthesis pipeline, we interpret each run as a strategy-synthesis outcome. The LLM may either output a candidate strategy or
fail to produce a valid one. If a candidate is produced, NatATL verification is used to decide whether the strategy is a certified winning strategy. In addition, we use an ATL pre-filter as a feasibility signal. The ATL pre-filter is not a replacement for NatATL verification. ATL checks whether a winning strategy exists in the unconstrained ATL sense. Since NatATL
restricts admissible strategies to bounded natural strategies, a negative ATL result rules out NatATL satisfiability, whereas a positive ATL result only
indicates that a winning strategy may exist before imposing the NatATL natural-strategy bound. Thus, ATL is used only to distinguish cases where failure is expected from cases where the LLM may have failed to find a strategy even though the unconstrained strategic objective is feasible. Algorithm~\ref{alg:llm-natatl} reports the resulting verification loop. The procedure first queries the LLM for a candidate strategy. If no valid strategy
is produced, the ATL pre-filter is used to distinguish true negatives from false negatives. If a candidate strategy is produced, the NatATL verifier checks it. Accepted candidates are classified as true positives. Rejected candidates are classified as false positives, and if the ATL check indicates that a winning strategy exists in the unconstrained sense, the verifier feedback is used to trigger a bounded retry. This classification protocol is used only for empirical evaluation. The soundness guarantee of the framework concerns accepted positive answers: a
strategy is returned as successful only if it has been certified by the NatATL verifier. All other labels describe failure modes of the LLM-guided synthesis process under the chosen retry and token budgets.

\begin{algorithm}[H]
    \caption{LLM-based Strategy Synthesis and Verification for NatATL}
    \footnotesize
    \label{alg:llm-natatl}
    \begin{algorithmic}[1]
        \Require MAS \(M\), NatATL formula \(\varphi\), LLM \(Q\)
        \Ensure Prediction \(P \in \{\mathsf{TP},\mathsf{TN},\mathsf{FP},\mathsf{FN}\}\), candidate strategy \(S\)

        \State \(\mathit{resp} \gets Q.\mathsf{ask}(M,\varphi)\)
        \State \(\mathit{feedback} \gets \bot\)

        \If{\(\mathit{resp}\) does not contain a valid candidate strategy}
            \State \((\mathit{atl}, \_, \_) \gets \mathsf{verify}(M,\bot)\)
            \If{\(\neg \mathit{atl}\)}
                \State \Return \((\mathsf{TN}, \bot)\)
            \Else
                \State \(\mathit{feedback} \gets\) ``A strategy may exist; generate a valid JSON strategy.''
                \State \(\mathit{resp} \gets Q.\mathsf{retry}(\mathit{feedback})\)
            \EndIf
        \EndIf

        \If{\(\mathit{resp}\) contains a valid candidate strategy}
            \State \((\mathit{atl}, \mathit{nat}, S) \gets \mathsf{verify}(M,\mathit{resp})\)

            \If{\(\mathit{nat}\)}
                \State \Return \((\mathsf{TP}, S)\)
            \ElsIf{\(\neg \mathit{atl}\)}
                \State \Return \((\mathsf{FP}, S)\)
            \Else
                \State \(\mathit{feedback} \gets\) verifier diagnostic for \(S\)
                \State \(\mathit{resp} \gets Q.\mathsf{retry}(\mathit{feedback})\)

                \If{\(\mathit{resp}\) contains a valid candidate strategy}
                    \State \((\_, \mathit{nat}, S) \gets \mathsf{verify}(M,\mathit{resp})\)
                    \If{\(\mathit{nat}\)}
                        \State \Return \((\mathsf{TP}, S)\)
                    \Else
                        \State \Return \((\mathsf{FP}, S)\)
                    \EndIf
                \Else
                    \State \Return \((\mathsf{FN}, \bot)\)
                \EndIf
            \EndIf
        \EndIf

        \State \Return \((\mathsf{FN}, \bot)\)
    \end{algorithmic}
\end{algorithm}

%% file: sections/05-Evaluation.tex
\section{Experiments}
\label{sec:experiments}


\begin{table}[h]
\centering
\small
\setlength{\tabcolsep}{8pt}
\renewcommand{\arraystretch}{1.2}
\begin{tabular}{c c c c}
\hline
\textbf{LLM} & \textbf{NatATL} & \textbf{ATL} & \textbf{Label} \\
\hline
false & *     & false & TN \\
false & *     & true  & FN \\
true  & false & false & FP--halt \\
true  & false & true  & FP--retry \\
true  & true  & *     & TP \\
\hline
\end{tabular}
\caption{Outcome matrix for the LLM-driven strategy synthesis tool.}
\label{tab:outcome-matrix}
\end{table}

To avoid unnecessary exhaustive exploration of the NatATL strategy space, our pipeline integrates an ATL pre-filtering phase before invoking full NatATL verification \cite{aruta2025s4h}. The rationale is that NatATL strategy synthesis is computationally expensive and becomes pointless when no winning strategy exists. While ATL does not synthesize strategies, it can efficiently determine whether a solution does not exist for the given model and objective. Leveraging ATL as a preliminary feasibility oracle therefore allows us to prune worst-case instances early, preventing the LLM from wasting computation on strategies that are guaranteed to fail under the fixed bound. Moreover, ATL checking is already embedded in real time within \textsc{VITAMIN}, and thus introduces negligible computation time overhead compared to NatATL verification. This makes the pre-filter not only a scalability optimization, but also a key component for improving training reliability: it provides structured signals that help distinguish genuine negative cases from bound-related failures. Table~\ref{tab:outcome-matrix} reports the possible outcomes of our pipeline. If the LLM does not output any strategy, the model-checking step is not applicable ("*") and we use the ATL checker to distinguish: (i) \textbf{TN} when ATL=false (no winning strategy exists), and (ii) \textbf{FN} when ATL=true (a winning strategy exists but the LLM failed to produce one; an exhaustive NatATL search may be required). If the LLM outputs a strategy but the NatATL model checker rejects it (NatATL=false), we classify the candidate as a \textbf{false positive} and use the ATL checker to decide whether to halt (\textbf{FP-halt}, ATL=false) or to run an additional refinement round (\textbf{FP-retry}, ATL=true). Finally, if the NatATL model checker validates the strategy (NatATL=true), the outcome is a \textbf{TP} and the ATL checker is unnecessary ("*").

\begin{figure}
    \centering
    \includegraphics[width=0.8\linewidth]{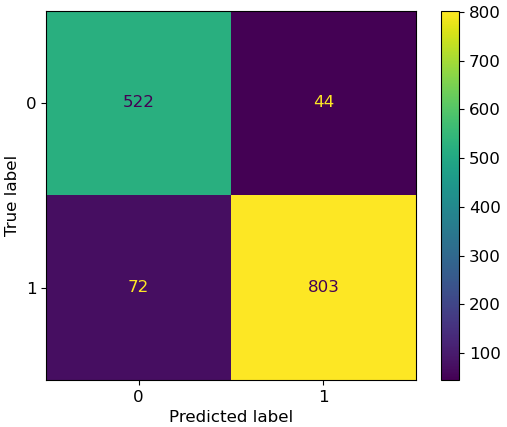}
    \caption{Confusion Matrix.}
    \label{fig:conf_mat}
\end{figure}

\begin{figure}
    \centering
    \includegraphics[width=1\linewidth]{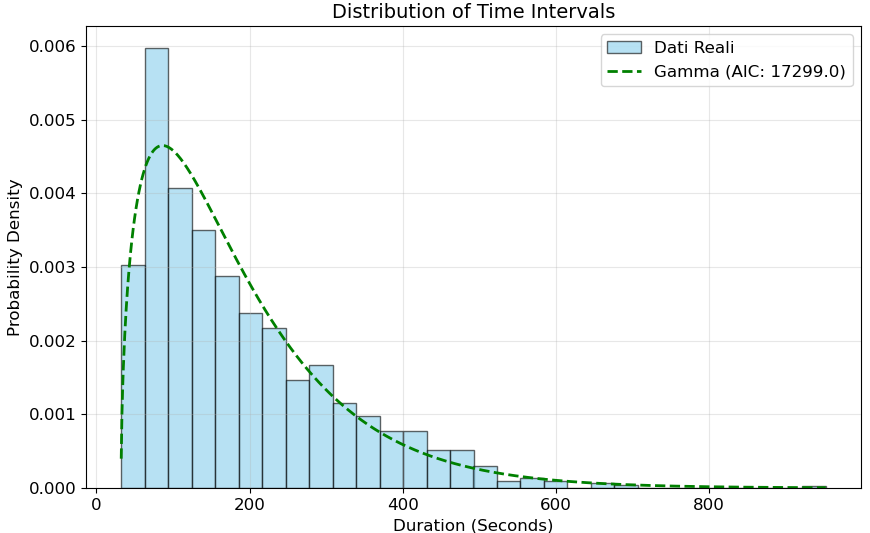}
    \caption{Time distribution in seconds.}
    \label{fig:time_dist}
\end{figure}

\paragraph{Results.} \label{sect:results}
After the strategy synthesis, we interpreted the results as a classification problem by labeling each sample as \emph{True Positive}, \emph{True Negative}, \emph{False Positive} and \emph{False Negative}, based on the analysis pipeline as reported in the Outcome interpretation paragraph. 
From the confusion matrix (Figure~\ref{fig:conf_mat}) we see that the LLM synthesis is very good, $41$ instances over $1411$ are marked as false positive and $72$ as false negative, this means that LLM with thinking modalities could cope with MAS of small to medium complexities under NatATL formula constraints with excellent accuracy of $92\%$. The time distribution that the pipeline required for each sample is plotted in Figure~\ref{fig:time_dist}, it best fits a Gamma distribution of parameters $shape=1.53, loc=32.23, scale=101.72$. By inferencing with at least two LLM iterations, we explicitly let the model to better handle the \textit{true-false-true} outcome in table \ref{tab:outcome-matrix} (LLM outputs a candidate, NatATL model checking fails, while the ATL checker indicates that a winning strategy exists). This case is hard to improve with a single LLM iteration, because without a second iteration the system cannot provide actionable feedback of the form: \emph{"a strategy does not exist for the given bound $k$, but it may exist for larger bounds"}. Such feedback helps the LLM interpret the instance correctly without \emph{forcing} it to always output a strategy, i.e., without implicitly encouraging it to ignore the resource bound when assessing feasibility. We, finally, need to mention the fact that not all samples are decidable under our framework because of the inherent limitation of our experimental setup, we had to put a cap on the maximum token number of $8192$, this makes that not all samples become decidable in the second query (or even in the first) to LLM giving a truncated output. Those cases are marked as \textit{ParsingError} and do not contribute to the performance metrics. Those \textit{ParsingError} account for $51$ samples and are not included into the final dataset samples.

\begin{table}[t]
\centering
\small
\renewcommand{\arraystretch}{0.9}
\setlength{\tabcolsep}{5pt}
\begin{tabular}{l r r r r r}
\hline
\textbf{Tool} &
\textbf{\#States} &
\textbf{$|C|$} &
\textbf{$k_{\max}$} &
\textbf{Verification time (s)} \\
\hline

NatSTV & 6   & 2 & 6  & $\approx0.8 $\\
NatSTV & 8   & 2 & 8  & $\approx12.4 $\\
NatSTV & 10  & 3 & 8  & $\approx95.7$\\
NatSTV & 12  & 3 & 10 & $\approx312.5$\\
NatSTV & 14  & 3 & 10 & timeout\\

\hline

Our Tool & 6  & 2 & 3 & $\approx70  $\\
Our Tool & 10 & 2 & 5 & $\approx80  $\\
Our Tool & 20 & 5 & 10 & $\approx 120 $\\
Our Tool & 50 & 11 & 100 &$\approx   200$\\

\hline
\end{tabular}
\caption{Scalability comparison between existing tools.} 
\label{tab:comparison}
\end{table}
Table~\ref{tab:comparison} highlights the radically different scalability behavior of current validated tools at the nowadays state of art (\cite{kaminski2025natstv}) and our approach when the size of the model, the coalition, and the complexity bound grow. NatSTV exhibits a clear exponential blow-up already at very small scales. While verification remains fast for models with up to 6 states, performance rapidly deteriorates as soon as the number of states and coalition agents increases. In particular, moving from 8 to 12 states causes an increase of more than one order of magnitude in runtime, ultimately leading to a timeout at 14 states. This confirms that NatSTV suffers heavily from the symbolic encoding of strategy choices combined with bounded complexity, which quickly saturates the underlying verification engine. In contrast, our tool maintains stable and predictable performance even under significantly larger configurations. Although the absolute runtime is higher for very small instances (due to explicit strategy reasoning rather than symbolic shortcuts), the growth remains smooth and controlled as both the coalition size and $k_{\max}$ increase. Notably, our approach successfully handles models with up to 50 states, coalitions of size 11, and bounds as high as $k=100$, remaining well within practical verification time. This demonstrates that the bottleneck affecting NatSTV and the other proposed tools in literature is effectively mitigated in our framework, which scales with the structure of the strategy space rather than collapsing under symbolic explosion. 

\paragraph{Reproducibility.}
We provide the source code and data used for the reproducibility of the experiments, you can download them at: \cite{SoftwareRep}.

%% file: sections/06-Conclusions.tex
\section{Conclusions}
We introduced a neuro-symbolic framework for certified strategy synthesis in NatATL, where LLMs act as strategy-generation oracles that propose natural strategies directly from a multi-agent model and a NatATL specification. Rather than replacing formal verification, the LLM is embedded in a generate-and-certify pipeline: candidate strategies are accepted only when formally validated by the NatATL model checker. This design preserves soundness while avoiding, in successful cases, the exhaustive enumeration of bounded natural strategies. We focused on NatATL because its bounded-rationality assumptions naturally capture the implementability constraints of real-world agents. Natural strategies are required to be simple, interpretable, and bounded in syntactic complexity, making them well aligned with practical strategic reasoning. This perspective also provides a natural interface with LLM-based synthesis, where token budgets and context-window limitations impose their own form of bounded reasoning. To support systematic evaluation, we introduced the first NatATL strategy-synthesis dataset, consisting of 4,211 instances of multi-agent games, NatATL specifications, and expert-validated natural strategies. The dataset provides a reusable benchmark for evaluating strategy-generation oracles and for developing future learning-based approaches to natural strategic reasoning. Our experiments show that an open-weight Qwen3-32B model, coupled with formal certification, can effectively synthesize natural strategies for small- to medium-sized multi-agent systems. On the evaluated benchmark split, the proposed pipeline achieves 92\% accuracy on strategy-synthesis outcomes, while every accepted positive result is backed by verifier certification. The results also show a favorable scalability profile compared with exhaustive natural-strategy verification, with our tool handling configurations up to 50 states, coalitions of 11 agents, and complexity bounds up to \(k=100\). At the same time, our study highlights practical limitations of LLM-guided verification. The maximum context length constrains the size of the serialized multi-agent models that can be processed by the LLM, and GPU memory limits the scale of inference for larger instances. These limitations suggest several directions for future work, including more compact model encodings, verifier-guided prompt compression, fine-tuning or distillation for NatATL-specific synthesis, and broader comparisons across open-weight and proprietary LLMs. Overall, our results indicate that LLM-guided synthesis combined with formal certification is a promising direction for scaling natural strategic reasoning without sacrificing correctness guarantees.

\paragraph{Future Works.}
This seminal study represents a springboard for various directions to explore in the future. The first task we have identified is the creation of a dataset and a format to standardize a benchmark for further studies on MAS synthesis. This would allow us to measure the progress of the field, making scientific works comparable. 
Furthermore, reasoning methods have recently been introduced in small-scale LLM programs; we expect exponential improvements over time, so a standardized benchmark could highlight these improvements.
Another promising direction is fine-tuning through distillation of largest LLM in which a basic LLM is fine-tuned specifically for model synthesis. Finally, a specific study to test the strategic capabilities of different LLM models, both open source and proprietary, would be interesting, in order to create a line of research to improve the strategic capabilities of future LLMs.